\documentclass[aps,notitlepage,a4paper,superscriptaddress,12pt]{article}

\usepackage{multicol}
\usepackage[utf8]{inputenc}
\usepackage[dvips]{color}
\usepackage[toc]{appendix}
\usepackage[hang,flushmargin]{footmisc} 
\usepackage{titlesec,titletoc,ntheorem-hyper,authblk,abstract,latexsym,microtype,booktabs,amsmath,hyperref,cleveref,fancyhdr,wrapfig,array,amsfonts, amssymb,geometry,appendix,cite, secdot,tensor,multirow,subcaption,cancel}
\usepackage{xcolor}

\newcommand{\be}{\begin{equation}}
\newcommand{\ee}{\end{equation}}
\newcommand{\bea}{\begin{eqnarray}}
\newcommand{\eea}{\end{eqnarray}}

\usepackage{graphicx,lipsum, cuted}
\usepackage{caption}
\usepackage{amsmath}
\usepackage{amsfonts}
\usepackage{amssymb}
\usepackage{stackengine}
\usepackage{mathtools}
\numberwithin{equation}{section}
\usepackage{titlesec}
\usepackage{sectsty}
\sectionfont{\bfseries\Large\raggedright}

\numberwithin{subcase}{case}
\usepackage{framed}
\allowdisplaybreaks
\usepackage[nottoc]{tocbibind}
\usepackage[free-standing-units]{siunitx}
\usepackage{helvet}
\usepackage{url}
\usepackage{titletoc}
\usepackage{blindtext}
\usepackage[gen]{eurosym}

\title{\textbf{Quantum tunneling driven by quintessence and the role of GUP}}
\author{Sauvik Sen}
 \affil{Department of Physics, Shiv Nadar Institution of Eminence,\\NH-91,  Gautam Buddha Nagar, 201314, Uttar Pradesh, India}
 
\affil{\textit{sauviksen.physics@gmail.com}}
\date{\today}

\begin{document}

\maketitle


\begin{abstract}
In this paper, we studied quantum tunneling of massless and massive particles pertaining to a Schwarzschild black hole in a quintessence background, and explored the consequences emerging from a generalized uncertainty principle (GUP). For the quintessence scenario, we considered two specific cases of $w$, which is the ratio of the pressure and energy density, namely $w=-1/3$ and $w=-2/3$. For the GUP, we used a modified Schwarzschild metric and employed a unique choice of contour integration to compute the tunneling amplitudes. An analysis and comparative study of the respective temperature profiles has been made. The energy emission rate has also been analysed.
\end{abstract} 

{\textbf{Keywords:} Quantum tunneling, black hole thermodynamics,  quintessence, GUP}

\maketitle

\section{Introduction}\label{sec:level1}
Albert Einstein's geometric theory of gravitation, namely general relativity, has been an area of intense exploration for more than a century. The subject of black holes is one of the most intriguing outcomes of the theory \cite{MTW,Carroll,PaddyBook}. The famous no-hair theorem states that black holes can be distinguished by three fundamental parameters, namely, the mass, charge, and angular momentum. The Schwarzschild case corresponds to one of the simplest solutions for a massive, non-rotating, spherically symmetric uncharged black hole. On the other hand, a black hole with mass and electric charge, but no spin is formulated by the Reissner-Nordstr\"{o}m solution. The Kerr black hole describes a generalization of the Schwarzschild metric wherein the spin is taken into account. Such a rotating and uncharged black hole goes to the Schwarzschild limit when the spin approaches zero. The Kerr-Newman solution can be considered to be the most general formulation that deals with a charged and rotating black hole. By applying the semiclassical approach of quantum field theory, Hawking \cite{Hawking1, Hawking2}  described how a black hole could emit radiation and eventually undergo a loss in its mass. As he explained, this was because of the possibility of pair production taking place near the event horizon caused by quantum fluctuations. 
Hawking took a quantum field theoretic approach in the context of a curved background to present his results, although he made no direct attempt to quantize the black hole itself.

Subsequently, several notable works have been reported with regard to the understanding of black hole's emission of radiation and the tied-up issue of tunneling. These include, in the main, the quantum tunneling approach of Parikh and Wilczek \cite{Parikh1} who applied the techniques of WKB approximation to evaluate the tunneling amplitude and utilize the radial null geodesic when dealing with the massless particles, and the Hamilton-Jacobi method of Angheben et. al. \cite{AnghebenJHEP}, alongside related papers like in \cite{Vagenas2, Majhi2, Nadalini, Chakraborty, Mann1}. Furthermore, tunneling of massive particles was demonstrated in models proposed in \cite{Zhang_NPB, Hu_MPLA, Deng} along with tunneling of massive fermions \cite{Mann2, Majhi1}. Other works addressed tunneling arising as a consequence of the first law of thermodynamics \cite{Sarkar1}, tunneling caused by a charged massive particle from Reissner-Nordstr\"om-de Sitter black hole interacting with a global monopole \cite{Jiang}, Hawking radiation from a $(d+1)$ - dimensional anti-de Sitter black hole \cite{Hemming}, the possibility of Unruh radiation where tunneling could also be estimated with the help of WKB approximation \cite{deGill}, estimating Planck scale corrections as done in \cite{Vagenas1}, analogue Hawking radiation in Weyl semimetal \cite{Sen_JMP}, and also in a two-level $PT$ symmetric system \cite{Sen_Entropy}, and tunneling happening in a BTZ black hole \cite{Sen_IJMPA, Medved1}. Tunneling of charged and uncharged particles across the horizon of a dilaton-axion black hole has been studied in \cite{Soumitra}. For a review of an alternate mode of tunneling in contrast to conventional Hawking radiation, see \cite{Vanzo_Review}. For tunneling in Kiselev like AdS spacetime in $f(R,T)$ gravity we refer to \cite{Ali1}. \cite{Ali2} describes a similar analysis of tunneling in a modified Schwarzschild–Rindler black hole while \cite{Ali3} is concerned with bosonic tunneling for a dyonic black hole surrounded by a perfect fluid. Finally, the case of tunneling in Einstein–Gauss–Bonnet black hole is analyzed in \cite{Ali4,Ali5}.

The simplest description of a black hole manifold is provided by the
Schwarzschild metric which reads
\begin{equation}\label{Schwarzschild}
    ds^2 = -f(r) dt^2 + f(r)^{-1} dr^2 + r^2 d\theta^2 + r^2 \sin^2\theta\,d\phi^2,
\end{equation}    
where
\begin{equation}\label{f(r)}
       f(r) = 1-\frac{2M}{r},
\end{equation}
and $r$ is the radial distance from the center of the black hole, and $M$ is the mass of the black hole. Throughout this work we shall adopt units $c=G=\hbar=k_B=1$, where $c$ is the speed of light in vacuum, $G$ is the Gravitational constant, $\hbar$ is the reduced Planck constant, and $k_B$ is the Boltzmann constant.

Our intention in this work is to study the tunneling effect for a Schwarzschild black hole in a background of quintessence \cite{Peeb1, Peeb2, Wett}. The latter is a time-varying, spatially-inhomogeneous, negative pressure component of the cosmic fluid \cite{Cald, Zla, WeinbergCosmo} and dynamic in character \cite{Linder,Nozari_Hajebrahimi}. A familiar example of quintessence is the energy of a scalar field which slowly evolves down its potential: the slowness ensures that the potential energy density exceeds the kinetic energy density. Note that our universe is expanding at an accelerated rate, and many experiments have supported this feature in the last few decades. A good review on the accelerated expansion is provided in \cite{RatraRMP}. One of the explanations for the expansion of the universe is due to the existence of a negative pressure omnipresent elusive material called dark energy. In this regard, the cosmological constant ($\Lambda$) arising in Einstein's equations acts as the most promising candidate of dark energy. Another strong contender for the role of dark energy is quintessence. As pointed out by Weinberg \cite{WeinbergCosmo}, the phenomenon of quintessence corresponds to the time-varying vacuum energy. Like all other alternative models, quintessence is essentially a scalar field coupled with gravity, which is dynamic and spatially inhomogeneous \cite{Nozari_Hajebrahimi}. This scheme is an effort to replace the dormant cosmological constant with a dynamic negative pressure component. The minimally coupled scalar field has a potential that is inversely proportional to the field's nature. This potential decreases when the field evolves with time and the field value increases \cite{Fischler_JHEP}.

The Lagrangian of a scalar field minimally coupled to gravity has the form
\begin{equation}
    \mathcal{L} = \frac{1}{2}\partial_\mu \phi \, \partial^\mu \phi - V(\phi)
\end{equation}
where $\phi$ is the scalar field and $V(\phi)$ is the scalar potential. The field $\phi$ experiences the gravity indirectly through the curvature of background spacetime, and the self-interaction is due to the scalar potential $V(\phi)$ \cite{Linder}. Corresponding to the above $ \mathcal{L}$, the energy-momentum tensor is given by 
\begin{equation}
    T_{\mu\nu}= \frac{2}{\sqrt{-g}} \frac{\delta(\sqrt{-g}\mathcal{L})}{\delta g^{\mu\nu}}
\end{equation}
where $g$ is the determinant of the metric tensor $g_{\mu\nu}$.

For a homogeneous and isotropic spacetime, the corresponding pressure ($\mathtt{p}$) and energy density ($\rho$) are 
\begin{subequations}
\begin{eqnarray}
\mathtt{p}&=&\frac{\dot{\phi}^2}{2}-V(\phi)-\frac{1}{6}(\nabla \phi)^2 \\
\rho &=& \frac{\dot{\phi}^2}{2}+V(\phi)+\frac{1}{2}(\nabla \phi)^2    
\end{eqnarray}    
\end{subequations}
where we have neglected the spatial gradient terms for both pressure and density because late time acceleration needs a very light scalar field, and the Compton wavelength of the field is of the order of (or larger) the Hubble scale, having a spatially smooth field within the Hubble scale \cite{Linder}. We can compute the usual Friedmann equations to solve for the expansion history of the scale factor $a(t)$. The relevant equation of state ratio is defined as  
\begin{equation}\label{EoS}
    w=\frac{\mathtt{p}}{\rho} = \frac{\displaystyle\frac{\dot{\phi}^2}{2}-V(\phi)}{\displaystyle\frac{\dot{\phi}^2}{2}+V(\phi)} \equiv \frac{K-V}{K+V}
\end{equation}

The kinetic energy term ($K$) and the potential energy term ($V$) can be written in terms of pressure and energy density 
\begin{subequations}
    \begin{eqnarray}
        V &=& \frac{\rho-\mathtt{p}}{2} = \frac{\rho}{2}(1-w) \\
        K &=& \frac{\rho+\mathtt{p}}{2} = \frac{\rho}{2}(1+w)
    \end{eqnarray}
\end{subequations}
As such, the scalar fields at any epoch is guided by one of the four possibilities namely, fast, slow, steady, or oscillatory \cite{Linder} type.

The cosmological constant corresponds to $w=-1$, whereas the universe has matter domination when $w=0$, and radiation dominance when $w=-1/3$ \cite{SusskindJHEP}. The observational bound on the cosmological constant is actually $-1<w\leq -2/3$. Quintessence is therefore in the range $-1 < w < -1/3$. The upper bound of quintessence acts as a phase transition border. In the range $-1 < w < -1/3$ the universe accelerates, for $w>-1/3$ the universe is decelerating, and at the boundary $w=-1/3$ the universe is expanding at a constant rate. Considering these possible ranges of $w$, we will focus on the boundary points of the most promising range for quintessence, i.e., $(-2/3,-1/3)$. Sometime ago, Kiselev \cite{Kiselev} determined the density for the quintessence matter in a Schwarzschild background, while GUP corrections were calculated in \cite{Nozari_Haje_Sag}. The thermodynamics of the Schwarzschild black hole in the context of quintessence in gravity’s rainbow was studied in \cite{Hamil}.

In the quintessence case, the function $f(r)$ in (\ref{Schwarzschild}) is replaced by a modified function $f^{(q)}(r)$ as given by \cite{Kiselev}
\begin{equation}\label{f^q(r)}
    f^{(q)}(r) = 1-\frac{2M}{r}-\frac{\alpha}{r^{3w+1}},
\end{equation}
where $\alpha$ is the constant of normalization, and $w$ is the ratio of pressure and energy density arising from the equation of state. We refer to $r_{in}$ as the radius of the event horizon when the particle is produced just inside it, while $r_{out}$ stands for the radius of the event horizon when the particle has tunneled out of the black hole. The region between $r_{in}$ and $r_{out}$ serves as the potential barrier that the tunneling particles have to penetrate. The horizon limits can be computed by setting $f^{(q)}(r)=0$. For $w=-1/3$, it gives for $r_{in}$ and $r_{out}$  the respective values 
\begin{eqnarray}
  &&  r_{in}^{(w=-1/3)} = \frac{2M}{1-\alpha}, \label{rin_13}\\
  &&  r_{out}^{(w=-1/3)} = \frac{2(M-\Omega)}{1-\alpha}, \label{rout_13}
\end{eqnarray}
while for $w=-2/3$ the corresponding ones are
\begin{eqnarray}
 && r_{in}^{(w=-2/3)} =  \frac{1+\sqrt{1-8\alpha M}}{2\alpha}, \label{rin_23}\\
 && r_{out}^{(w=-2/3)} = \frac{1+\sqrt{1-8\alpha(M-\Omega)}}{2\alpha}.\label{rout_23}
\end{eqnarray}
We note that $r_{out}$ in (\ref{rout_13}) and (\ref{rout_23}) are the radii of the horizon when a particle of mass $\Omega$ emits out.

In different models of quantum gravity (QG), it is widely accepted that the existence of a nonzero minimal length is imperative \cite{Mead_PhysRev}. The common feature of loop quantum gravity (LQG), string theory, and noncommutative geometry \cite{Das1, Kato, Hossenfelder} is the occurrence of a Planck scale order minimal length wherein the famous Heisenberg uncertainty principle (HUP) is modified to get a new generalized uncertainty principle (GUP). A new QG theory like double special relativity (DSR) deals with a UV cutoff for quantum field theory (QFT) by implementing a minimal length and thereby putting an upper bound on the maximum permissible momentum \cite{Hannawi,Nozari_PRD}. The fundamental length in the case of black holes helps to address the information loss paradox. GUP induced cutoff energy resolves the vacuum energy divergence issue of QFT.

In black hole physics, it is common to introduce GUP by deforming the mass of the black hole \cite{Anacleto3}. In cosmology, GUP has been known to show acceleration of the universe in the early days but in the later stages has been causing deceleration \cite{Zeynali1, FaragAli2}. For the case of the positive cosmological constant, the commutative and GUP cases predict infinite-size universes, which show late-time acceleration behavior in accordance with current observations \cite{Zeynali2}.
The concepts of minimal length and cutoff energy are used to tackle the dark energy problem \cite{Vakili, Kouwn, Majumdar_AdvHEP}.

Against the background of an earlier work \cite{Eslamzadeh_Nozari} on the tunneling of massless and massive particles corresponding to a
quantum deformed Schwarzschild black hole surrounded by quintessence, the primary aim of the present study is to extend the well-known tunneling approach to investigate exotic quintessence and GUP correction behavior on a Schwarzschild black hole. We will address tunneling in a semi-classical WKB approximation for the two explicit cases corresponding to 
(i) the case of a black hole in a background quintessence field, and (ii) for a deformed Schwarzschild metric in the framework of a generalized uncertainty principle (GUP). In (i) we pursue a classical treatment using a scalar field coupled to gravity and in (ii) we consider the quantum effects as a consequence of the GUP. We might remark here that in contrast to the conventional Heisenberg principle, which speaks of an inequality in the phase space between the pair of complementary measurable variables like the position ($x$) and momentum ($p$), namely,
$\Delta x \Delta p \geq \frac{\hbar}{2}$,
there are several accounts of its extended version in the context of sub-Planckian black holes, where the particle interacts gravitationally with the photon. Up to second order in $\Delta p$, a plausible modification of the Heisenberg inequality was given in \cite{SauryaDas1,SharuyaDas3}.
\begin{equation}\label{GUP}
    \Delta x\Delta p \geq \frac{\hbar}{2}\left(1 - \frac{\beta l_p}{\hbar}\Delta p + \frac{\beta l_p^2}{\hbar^2} (\Delta p)^2     \right),
\end{equation}
where $\beta$ is a model-dependent deformation parameter (see \cite{Kanazawa} and references therein), and $l_p$ is the Planck length (we take $\hbar=1$ in our work). Similar approaches have also been pursued in \cite{Kempf, Bagchi, Anacleto1,Anacaleto2,Anacleto3}. In (\ref{GUP}), both the linear and quadratic terms in $\Delta p$ are retained. The above form takes into account plausible corrections to the Heisenberg inequality in a way that any localization in space is prevented. A comparative study of two GUPs with a minimal length/maximal momentum and the other with all natural cutoffs has been done in the context of black hole thermodynamics \cite{Nozari_JHEP}. The role of gravitational waves to restrict the GUP parameter was analysed in \cite{Feng2}. The discrete character of space was studied through the use of the GUP in \cite{FaragAli1,ShauryaDas2}. For further insightful discussions on GUP see \cite{ShauryaDas4,FaragAli3}.

Expanding the new deformed mass $\mathcal{M}(r)$ in terms of the deformation parameter $\beta$ gives up to its first order \cite{Anacleto3} 
\begin{equation}\label{eq3.2}
    \mathcal{M}=M\left(1-\frac{2\beta}{M}+\frac{4\beta}{M^2}\right).
\end{equation}
As a consequence, the function $f(r)$ of (\ref{Schwarzschild}) converts accordingly to $f^{(g)}(r)$ gets transformed to
\begin{equation}\label{f^g(r)}
   f^{(g)}(r) = 1-\frac{2\mathcal{M}}{r}=1-\frac{2M}{r}\left(1-\frac{2\beta}{M}+\frac{4\beta}{M^2}\right),
\end{equation}
The horizon limits are now to be extracted by solving $f^{(g)}(r)=0$. The corresponding $r_{in}$ and $r_{out}$ turn out to be respectively 
\begin{eqnarray}\label{r^g}
  &&  r^{(g)}_{in} = 2M\left(1-\frac{2\beta}{M}+\frac{4\beta}{M^2}\right), \\
  &&  r^{(g)}_{out} = 2(M-\Omega)\left(1-\frac{2\beta}{(M-\Omega)}+\frac{4\beta}{(M-\Omega)^2}\right).
\end{eqnarray}
where the dimensionless parameter $\beta$ scales with $l_p$. Due to our choice of units $l_p=\sqrt{\frac{\hbar G}{c^3}}=1$. 

Our work is arranged as follows.
Section \ref{sec:label3} is devoted to the approach of radial null geodesic to derive a tractable expression for the tunneling probability. In Section \ref{sec:label4} we evaluate the tunneling probabilities for massless particles corresponding to both quintessence and GUP cases. Section \ref{sec:label5} is concerned with similar calculations for the massive particles. In Section \ref{sec:label6}, we address the black hole thermodynamics from the perspective of different approaches. Finally, we present a summary and conclusions of our results in Section \ref{Conclusion}. Details on the solutions to the contour integrals are presented in the appendices \ref{Appendix-A} and \ref{Appendix-B}. 

\section{Approach of radial null geodesic}\label{sec:label3}

Following Parikh and Wilzcek \cite{Parikh1}, we make the following translatory transformation where the time coordinate $t$ in the Schwarzschild metric is shifted to Painlev\'e time $T$ 
\begin{equation}\label{eq4.1}
    T=t+\xi(r),
\end{equation}
where $\xi(r)$ is an arbitrary function of $r$. This gives straightforwardly 
\begin{equation}\label{eq4.2}
    dt^2 = dT^2 + \xi'(r)^2 dr^2 - 2\xi'(r) dr dT,
\end{equation}
wherein a prime indicates a derivative with respect to $r$.
Substituting (\ref{eq4.2}) in modified (\ref{Schwarzschild}) produces 
\begin{multline}
   ds^2 = -f^{(i)}(r) dT^2 + \big( f^{(i)}(r)^{-1} - f^{(i)}(r)\xi'(r)^2\big)dr^2
   + 2f^{(i)}(r)\xi'(r)drdT\\
+r^2d\theta^2+r^2\sin^2\theta d\phi^2,  \quad
   i = q, g. \label{eq4.3}
\end{multline}
The use of Painlev\'e-Gullstrand (PG) \cite{Painleve,Gullstrand} coordinates enables us to remove the coordinate singularities that are present in the black hole's metric. Unlike curvature singularities that are true singularities (eg. $r=0$ for Schwarzschild metric), the coordinate singularities ($r=2M$ for Schwarzschild) arise because of the choice of an imperfect coordinate system. To tackle this problem, it is useful to employ PG coordinates, often referred to as the global rain coordinates \cite{TaylorWheeler}. The name comes from the coordinate system attached to a drop of rain falling into the black hole in its own frame of reference, instead of a global inertial time frame of Schwarzschild coordinates. This new time coordinate is necessary because this raindrop is able to cross the event horizon of the black hole without facing a discontinuity in its trajectory. This transformation makes the metric globally continuous except at $r=0$. Using the PG coordinates, the raindrops are now able to cross the event horizon without facing any discontinuity with this new time coordinate, which can be written in an exact differential form in terms of the Schwarzschild time $t$ and the radial coordinate $r$. A metric derived in terms of the PG coordinates ($T,r,\theta,\phi$), by transforming a valid metric originally given in Schwarzschild coordinates ($t,r,\theta,\phi$), is allowed. This is due to the principle of general covariance. Here $T$ is the new time coordinate called the Painlev\'e time. There are other global rain coordinates as well. Eddington-Finkelstein coordinates and Kruskal-Szerkes coordinates are to name a few. These PG coordinates are important in our current study to calculate the tunneling probability of the outgoing particles that cross the event horizon.

For the radial null geodesic, we set the coefficient of $dr^2$ to be unity, which implies 
\begin{equation}\label{eq4.5}
    \xi(r)=\pm\int\frac{\sqrt{1-f^{(i)}(r)}}{f^{(i)}(r)} dr.
\end{equation}
The metric is now furnished in PG coordinates, namely
\begin{equation}\label{eq4.6}
ds^2=-f^{(i)}(r) dT^2 + dr^2 + 2\sqrt{1-f^{(i)}(r)}dr dT + r^2 d\theta^2 + r^2 \sin^2 \theta d\phi^2.
\end{equation}
The constraint $\theta=\theta_0$ gives the radial null geodesic slice for a particular $\theta$ as guided by the differential equation 
\begin{equation}\label{eq4.7}
\dot{r}^2 + 2\sqrt{1-f^{(i)}(r)}\,\dot{r}-f^{(i)}(r)=0\,,   
\end{equation}
where $\dot{r}\equiv\frac{dr}{dT}$. As a result, we have for the radial velocity 
\begin{equation}\label{eq4.8}
  \dot{r}=\pm1-\sqrt{1-f^{(i)}(r)}\,,
\end{equation}
where the two signs correspond to the outgoing and ingoing particle tunneling out or into the horizon, respectively. Since we are interested in the tunneling of outgoing particles, in the following we will choose the positive sign in (\ref{eq4.8}). 

The action $\zeta$ of the outgoing particle is given in the standard way by the integral $\zeta = \int \mathcal{L}\,dT$, where we know from classical mechanics that the Lagrangian $\mathcal{L}$ is linked to the Hamiltonian $H$ by means of the Legendre transformation $\mathcal{L} = p_r \dot{r} - H$, where $(p_r,r)$ are the conjugate pair of momentum and position variables in the radial direction. The time interval during which the emission of the tunneling takes place is very small, and hence, one can neglect the second term of the action. $\zeta$ therefore assumes the form 
\begin{equation}\label{zeta}
    \zeta = \int p_r\, dr,
\end{equation}
and the tunneling probability of the outgoing emitted particles is given by \cite{Parikh2, Parikh3}
\begin{equation}\label{tunnel}
    \mbox{Im}\, \zeta = \mbox{ Im} \int_{r_{in}}^{r_{out}} p_r \,dr = \mbox{ Im} \int_{r_{in}}^{r_{out}} \int_{0}^{p_r} dp_r\, dr.
\end{equation}

In (\ref{tunnel}), observe that, following emission of the tunneling particles, $r_{out}$ is actually smaller than $r_{in}$.
To proceed further, we note that in (\ref{tunnel}) Hamilton's equations are used to identify momentum in terms of the Hamiltonian $d p_r = \frac{dH}{v_{c}}$, were $v_{c}$ is the coordinate velocity, which for the massless case is the derivative of the radial component with respect to the Painlev\'e time ($v_{c}\equiv\dot{r} =  \frac{\partial r}{\partial T}$). The change in the Hamiltonian corresponds to the loss in mass of the black hole by a quantity $\Omega$ due to the emission of the particles, implying $dH \approx dM$, where $ dM \approx -d\Omega$ (see also \cite{Sen_IJMPA}). Thus changing the limits of the integration in (\ref{tunnel}) from $0$ to $\Omega$ the tunneling probability is given by \cite{Parikh1}
\begin{equation}\label{tunneling_probability}
 \Gamma = e^{-2\mbox{Im}\zeta},
\end{equation}
where $\mbox{Im}\, \zeta$ stands for the double integral
\begin{equation}\label{Omega}
    \mbox{Im}\, \zeta = -\mbox{Im} \int_{r_{in}}^{r_{out}} \int_{0}^{\Omega} \frac{d\Omega}{1-\sqrt{1-f^{(i)}(r)}} dr.
\end{equation}

Comparing with the Boltzmann factor one recognizes that tunneling probability can be cast as $\Gamma \sim e^{\Delta \mathsf{S_{BH}}}$, where $\mathsf{S_{BH}}$ is the change in the entropy of the black hole. It is useful to consider the WKB approximation here because it is valid only in the regions where the wavelength of the particle is much smaller than the distance it has to travel over which the potential varies rather slowly \cite{Sakurai}. The ansatz is taken to be of the form $\psi(r) \sim e^{i \,\zeta(r)}$ such that in the first order $\zeta(r)$ takes the form of the action $\int_r p(r) dr$. Due to the absence of a classical path, the particle is considered to tunnel across the imaginary part of the action. The tunneling probability for such a particle crossing the potential barrier (in this case, the event horizon) is then given by the form (\ref{tunneling_probability}) \cite{RShankar,ZettiliQM}. To estimate the tunneling probability as follows from (\ref{Omega}), we now seek to address the issues of quintessence and GUP.

\section{Tunneling of massless particles}\label{sec:label4}

First, we take up the case of $w=-\frac{1}{3}$. Here (\ref{f^q(r)}) becomes 
\begin{equation}\label{eq5.a.1.1}
    f^{(q)}(r) = 1 - \frac{2(M-\Omega)}{r}-\alpha,
\end{equation}
wherein we replaced $M$ by $M-\Omega$ because we are dealing with a particle which has just tunelled out of the horizon.

Evaluating the integral in (\ref{Omega}) according to the prescription described in Appendix \ref{Appendix-A}, the imaginary part of the action gets transformed to
\begin{equation}
    \mbox{Im} \zeta = \frac{2\pi}{(1-\alpha)^2} \left( M\Omega - \frac{\Omega^2}{2} \right).
\end{equation}
See Fig. \ref{contour1} for the determination of the contour integral. The tunneling probability acquires the form 
\begin{equation}
    \Gamma \sim e^{-\frac{4\pi}{(1-\alpha)^2}\left(M\Omega-\frac{\Omega^2}{2}\right)}.
\end{equation}

We see that $\Gamma$ is exponentially decaying, as can be expected of a quantum particle that tunnels through a forbidden barrier. When one suppresses $\alpha$, we get back the tunneling probability of a Schwarzschild black hole. Note that our result has a factor of $4\pi$ in the argument, which is half that of \cite{Parikh1}. The reason is that we have only dealt with the outgoing particle. Inclusion of the equal contribution from the ingoing antiparticle makes up for the factor $8\pi$ estimated in \cite{Parikh1}.

Next, for the case $w=-\frac{2}{3}$, $ f^{(q)}(r)$ takes the reduced form $  f^{(q)}(r) =
1 - \frac{2(M-\Omega)}{r}-\alpha r$. It gives the expression for the imaginary part of the action
\begin{equation}
    \mbox{Im}\, \zeta = \frac{\pi}{2} \left( \Omega + \frac{1}{4\alpha}\left(\sqrt{1-8\alpha(M-\Omega)}-\sqrt{1-8\alpha M}\right)\right),
\end{equation}
implying for the tunneling probability
\begin{equation}
    \Gamma \sim e^{-\pi \left( \Omega + \frac{1}{4\alpha}\left(\sqrt{1-8\alpha(M-\Omega)}-\sqrt{1-8\alpha M}\right)\right)}.
\end{equation}
$\Gamma$ is again a decaying function on $\Omega$. Expanding binomially, we find that for small $\alpha$, $\Gamma$ is independent of the mass $M$ of the black hole ($\Gamma \sim e^{-2\pi \Omega}$) in contrast to the $w=-1/3$ case where dependence on $M$ can be seen even for the small $\alpha$ approximation.

Turning to the GUP form of $ f^{(i)}(r)$, we substitute (\ref{f^g(r)}) in (\ref{Omega}) and solve the integral as we did for the previous case. Thus, we arrive at
\begin{equation}
    \mbox{Im}\, \zeta = \pi \left( (2M-4\beta)\Omega - \Omega^2 -8\beta \ln \left( \frac{|\Omega-M|}{M} \right)  \right).
\end{equation}
It yields the tunneling probability 
\begin{equation}\label{GUPmassless}
    \Gamma \sim e^{-2\pi \left( (2M-4\beta)\Omega - \Omega^2 -8\beta \ln \left( \frac{|\Omega-M|}{M} \right)  \right)}.
\end{equation}

We remark that by switching off the GUP correction, we get back the Schwarzschild case tunneling probability of \cite{Parikh1}. Also, rearranging (\ref{GUPmassless}) in the form $\Gamma \sim A \,e^{-4\pi\left(\left(M-2\beta\right)\Omega-\frac{\Omega^2}{2}\right)}$, where the scale factor $A\equiv(\frac{|\Omega-M|}{M})^{16\pi\beta}$ has approximately a unit value for $M>>\Omega$. This result shows that the GUP correction has a negligible effect on the tunneling behaviour as one recovers the Schwarzschild black hole's tunneling probability for a small $\beta$ correction.

\section{Tunneling of massive particles}\label{sec:label5}

For a particle with nonzero mass $m$, the Lagrangian of an affinely parametrized null or non-null geodesic is given by
\begin{equation}\label{eq5.2}
\mathcal{L}=\frac{m}{2}\left(\right. -f^{(i)}(r)t_{,\tau}^2 + f^{(i)}(r)^{-1} r_{,\tau}^2 + r^2 \theta_{,\tau}^2 +r^2 \sin^2\theta \phi_{,\tau}^2 \left.\right),
\end{equation}
 where we have used the comma notation for partial derivative $x_{,\tau}=\frac{\partial x}{\partial \tau}$ \cite{Narlikar}. Here $x \in \{t,r,\theta.\phi\}$ and  $\tau$ is the proper time. Invoking a similar condition like (\ref{eq4.5}) namely, $\chi(r)=\int\frac{\pm\sqrt{1-f^{(i)}(r)}}{f^{(i)}(r)}dr$ to ensure a unit coefficient for $r_{,\tau}^2$, the Painlev\'e time $T$ being defined to be $T=t+\chi(r)$, the Lagrangian reads \cite{Eslamzadeh_Nozari,Miao_EPL,Miao_GRG}
\begin{equation}\label{eq5.7}
    \mathcal{L} = \frac{m}{2} \Big( \Big. -f^{(i)}(r) T_{,\tau}^2 +r_{,\tau}^2 +2\sqrt{1-2f^{(i)}(r)}\,r_{,\tau}T_{,\tau} + r^2\theta_{,\tau}^2+r^2\sin^2\theta\phi_{,\tau}^2 \Big. \Big).
\end{equation} 
The absence of an explicit $T$ coordinate in the Lagrangian indicates that the corresponding conjugate momentum $P_T$ is a constant, say $-\Omega$. That is
\begin{equation}\label{eq5.8}
    P_T= \frac{\partial \mathcal{L}}{\partial T_{,\tau}} = -\Omega.
\end{equation}
Solving equation (\ref{eq5.8}) we get
\begin{equation}\label{eq5.9}
    T_{,\tau} = \frac{\sqrt{1-f^{(i)}(r)}\,r_{,\tau}+\frac{\Omega}{m}}{f^{(i)}(r)}.
\end{equation}
The timelike condition, or more precisely, the four-velocity normalization follows from (\ref{eq5.7}) and corresponds to
\begin{equation}\label{eq5.6}
 -f^{(i)}(r) T_{,\tau}^2 +r_{,\tau}^2+2\sqrt{1-f^{(i)}(r)}r_{,\tau}T_{,\tau}
+r^2\theta_{,\tau}^2+r^2\sin^2\theta\phi_{,\tau}^2=-1. 
\end{equation}
For a massive particle moving along a radial timelike geodesic, the angular terms in (\ref{eq5.6}) drop out and one is left with the expression
\begin{equation}\label{eq5.10}
r_{,\tau} = \pm \frac{\Omega}{m}\sqrt{1-\frac{m^2}{\Omega^2}f^{(i)}(r)}.    \end{equation}
From (\ref{eq5.9}) one then obtains
\begin{equation}\label{eq5.11}
    T_{,\tau}=\pm\frac{\Omega}{m}\frac{\left(\sqrt{1-f^{(i)}(r)}\sqrt{1-\frac{m^2}{\Omega^2}f^{(i)}(r)}+1\right)}{f^{(i)}(r)}.
\end{equation}
In consequence, the coordinate velocity emerges from the following ratio
\begin{equation}\label{eq5.12}
    v_c \equiv \frac{r_{,\tau}}{T_{,\tau}} = \frac{\pm f^{(i)}(r)\sqrt{1-\frac{m^2}{\Omega^2}f^{(i)}(r)}}{1\pm\sqrt{1-f^{(i)}(r)}\sqrt{1-\frac{m^2}{\Omega^2}f^{(i)}(r)}}.
\end{equation}
Choosing the positive sign for $v_c$ due to the positivity of the momentum of the particle, the tunneling probability becomes
\begin{equation}\label{eq5.13}
    \mbox{Im}\zeta = 
     -\mbox{Im}  \int_{r_{out}}^{r_{in}} \int_m^\Omega \frac{1+\sqrt{1-f^{(i)}(r)}\sqrt{1-\frac{m^2}{\Omega^2}f^{(i)}(r)}}{f^{(i)}(r)\sqrt{1-\frac{m^2}{\Omega^2}f^{(i)}(r)}}\;d\Omega\;dr.
\end{equation}

Splitting the integrand into two parts
\begin{equation}\label{eq5.14}
    \mbox{Im}\zeta =\mbox{Im}\int_m^{\Omega} (I_1 + I_2)\; d\Omega,
\end{equation}
where
\begin{subequations}
\begin{eqnarray}
I_1 &=& \int_{r_{in}}^{r_{out}} \frac{1}{f^{(i)}(r)\sqrt{1-\frac{m^2}{\Omega^2}f^{(i)}(r)}}\;dr  ,  \label{eq5.15a}
\\
I_2 &=& \int_{r_{in}}^{r_{out}} \frac{\sqrt{1-f^{(i)}(r)}}{f^{(i)}(r)}\;dr ,  \label{eq5.15b}
\end{eqnarray}
\end{subequations}
we are in a position to examine the respective cases of quintessence and GUP that follow from $I_1$ and $I_2$.
Just like the massless case, we focus on the specific values of $w=-\frac{1}{3}$ and $w=-\frac{2}{3}$. 

Let us consider the $w=-1/3$ case. Here
\begin{equation}\label{eq5.a.1.3}
    I_1 = \int_{r_{in}}^{r_{out}} \frac{1}{f^{(q)}(r)\left(1-\frac{1}{2}\frac{m^2}{\Omega^2}f^{(q)}(r)\right)}\;dr.
\end{equation}
Since near the horizon $f^{(q)}(r)$ is very small, we carry out a binomial expansion for it in the vicinity of the event horizon to obtain for $I_1$
\begin{equation}
    I_1=\frac{i\pi}{(1-\alpha)^2}(M-\Omega),
\end{equation}
where the detailed evaluation of the integral has been shown in the Appendix \ref{Appendix-B}. 

Just like the $I_1$ integral, we also carry out a binomial approximation for $I_2$. $I_2$ is thus computed to be
\begin{equation}\label{eq5.a.1.8}
    I_2 = \frac{i\pi}{(1-\alpha)^2}(M-\Omega),
\end{equation}
and turns out to be equal to $I_1$. The detailed evaluation of the integral $I_2$ is shown in the Appendix (\ref{Appendix-B}).

The two integrals $I_1$ and $I_2$ combine to give
\begin{eqnarray}\label{eq5.a.1.9}
\mbox{Im}\,\zeta &=& \frac{2\pi}{(1-\alpha)^2}\int_m^\Omega (M-\Omega)\, d\Omega  \nonumber\\
&=& \frac{2\pi}{(1-\alpha)^2}\left(M\Omega-\frac{\Omega^2}{2}-mM+\frac{m^2}{2}\right),
\end{eqnarray}
implying for the tunneling probability, the result 
\begin{equation}\label{tunnelingQ1}
    \Gamma \sim e^{- \frac{4\pi}{(1-\alpha)^2}\left( M\Omega - \frac{\Omega^2}{2} -Mm +\frac{m^2}{2}        \right)},
\end{equation}
which is marginally larger than that determined in the massless case.

For the case of $w=-2/3$ we notice an important condition that places an upper bound on the normalization constant $\alpha\leq1/8M$ as a natural result for the positivity of the expression under the square root. This is true for the assumption that the mass of the black hole is much greater than the maximum possible energy of the emitted particle $(M>>\Omega)$ and also that the maximum and minimum energy of the emitted particle are of the same order $(\Omega \gtrsim m)$. We make a comment here. Such a stringent upper bound on $\alpha$ would occur for subsequently smaller values of $w$.

Similar to the previous case, we can split the radial integral into two parts, namely $I_1$ and $I_2$, which can then be solved according to Appendix \ref{Appendix-B}. Unlike the massless case, $I_1$ here will have four poles, namely
\begin{equation}\label{eq47}
   \frac{1\pm\sqrt{1-8\alpha(M-\Omega)}}{2\alpha}, \quad
  -\frac{(2\Omega^2-m^2)\pm\sqrt{m^4(1-8\alpha(M-\Omega))+4\Omega^2(\Omega^2-2m^2)}}{2m^2\alpha},
\end{equation}
On close inspection, we note that the location of one of the poles coincides with $r_{out}$ as given in (\ref{rout_23}). This is expected since it is the outgoing particle that faces the singularity. The other three poles lie in the unphysical region.

Solving the much involved integrals $I_1$ and $I_2$ one obtains
\begin{equation}\label{I1forw23}
    I_1 = \frac{i \pi \alpha m^2}{4} \left( 1 + \frac{1}{\sqrt{1-8\alpha(M-\Omega)}} \right),
\end{equation}
and
\begin{equation}\label{I2forw23}
    I_2 = \frac{i \pi}{4} \left( 1 + \frac{1}{\sqrt{1-8\alpha(M-\Omega)}} \right).
\end{equation}
The sum of both the above integrals acts as the final integration of the imaginary action, which has the final form
\begin{equation}
    \mbox{Im}\,\zeta = \frac{\pi}{2} \left( \frac{(1+\alpha m^2)}{2} \left\{ (\Omega-m) + \frac{1}{4\alpha}\left(\sqrt{1-8\alpha(M-\Omega)}-\sqrt{1-8\alpha(M-m)}\right)\right\} \right).
\end{equation}
The unequal contribution of $I_1$ and $I_2$ as opposed to the $w=-1/3$ case, is an indication that the necessary condition imposed on $\alpha$ is vital for the current case. This is evident from the fact that an extra coefficient of $\alpha m^2$ is present in $I_1$ in (\ref{I1forw23}).

The tunneling probability is then
\begin{equation}
    \Gamma \sim e^{-\pi\left( \frac{(1+\alpha m^2)}{2} \left\{ (\Omega-m) + \frac{1}{4\alpha}\left( \sqrt{1-8\alpha(M-\Omega)} - \sqrt{1-8\alpha(M-m)} \right) \right\} \right)}.
\end{equation}
In the above expression for $\Gamma$, carrying out a binomial expansion, we see that the two square root terms simplify to yield, to first order in $\alpha$, the argument value  $-\pi\left(\left(1+\alpha m^2\right)\left(\Omega-m\right)\right)$. Since the factor $\alpha m^2$ is extremely small, we can safely ignore it so that the argument further approximates to the form $\Gamma \sim e^{-\pi (\Omega-m)}$. It implies, for $\Omega > m$, $\Gamma$ to be larger in comparison with the massless case. 

To address the GUP case for the massive particle, we need to consider $f^{(g)}(r)$ according to (\ref{f^g(r)}). The integrals $I_1$ and $I_2$ are then evaluated. We find
\begin{eqnarray}
  &&  I_1 = i\pi (M-\Omega) \left( 1-\frac{2\beta}{M-\Omega} + \frac{4\beta}{(M-\Omega)^2}  \right), \\
&& I_2 = i\pi (M-\Omega) \left( 1-\frac{2\beta}{M-\Omega} + \frac{4\beta}{(M-\Omega)^2}  \right),
\end{eqnarray}
which gives for the imaginary action 
\begin{equation}
\mbox{Im}\, \zeta =\pi \left(  (2M-4\beta) (\Omega-m) - (\Omega^2-m^2)-8\beta \ln \left( \frac{|\Omega-M|}{|m-M|}  \right) \right).
\end{equation}

As a result, the following expression of the tunneling probability emerges 
\begin{equation}
    \Gamma \sim e^{-2\pi \left(  (2M-4\beta) (\Omega-m) - (\Omega^2-m^2)-8\beta \ln \left( \frac{|\Omega-M|}{|m-M|}  \right) \right)}.
\end{equation}
We at once see that the tunneling probability for the GUP case has a marginally higher value than its massless counterpart.

\section{Thermodynamical aspects and related issues}\label{sec:label6}

With the results obtained in the previous sections for the quintessence and GUP cases, we now turn to their impact on black hole thermodynamics. According to the first law of black hole thermodynamics \cite{Bardeen_thermodynamics, Wald_thermodynamics}, we have  

\begin{equation}
    \delta M = \frac{\kappa}{8\pi} dA + \omega\, dJ + \Phi\, dQ
\end{equation}
where $\delta M$ is the change in the energy/mass of the black hole, $\kappa$ is the surface gravity, $A$ is the surface area of the event horizon, $\omega$ is the angular velocity, $J$ is the angular momentum, $\Phi$ is the electrostatic potential and $Q$ is the electric charge. When one compares with the first law of thermodynamics, it is readily seen that $\frac{\kappa}{8\pi} dA \sim T dS$ in tune with Bekenstein's observation that the area of the black hole is directly proportional to the entropy \cite{Bekenstein1}. (Also, see \cite{Sen1,Sen2} for recent works on black hole area quantization.) Having already calculated the tunneling probability, we collate the exponentially decaying function with the Boltzmann factor. For a particle of mass having unit energy, the Hawking temperature can be readily identified to be \cite{Parikh1}
\begin{equation}
    |T_H| = \frac{1}{\Delta S_{BH}}
\end{equation}
where $\Delta S_{BH}$ is the change in the entropy of the black hole \cite{Bardeen_thermodynamics, Wald_thermodynamics}. We provide in Table \ref{Table-1} our results for the Hawking temperature $T_H$ for both massless and massive particles.

We also display in Fig. \ref{Fig1} all cases of $T_H$ from where it is clearly reflected that $T_H$ always assumes lower values for the massless case than what one has for the nonzero mass. Fig. \ref{Fig2} showcases the massless case for two values of $\alpha$, namely, $\alpha=0.25$ and $\alpha=0.1$, while Fig. \ref{Fig3} addresses the massive case for the same $\alpha$ values. Closer inspection reveals that the trend is not monotonic in all cases. For instance, as is evident from Fig. \ref{Fig2a}, the temperature is maximum for the GUP for a given mass of black hole, while for $w=-2/3$ it is minimum for the quintessence. This feature is disrupted in Fig. \ref{Fig2b} in which for $w=-2/3$ the temperature peaks for an intermediate range of $M$. One can draw similar conclusions for the massive case through Fig. \ref{Fig3a} and Fig. \ref{Fig3b}. It needs to be mentioned that the temperature profile for the $w=-2/3$ case has also been studied in \cite{Shahjalal, Eslamzadeh_Nozari}. In particular, \cite{Eslamzadeh_Nozari} considered the splitting of the black hole horizon into two parts, the outer horizon being termed as the cosmological horizon. At the critical value of $\alpha=1/8M$, a current flow took place from the black hole horizon to the cosmological horizon. Hawking-like radiation in the cosmological black hole solution for a modified gravity model has been studied in \cite{Saghafi1}. The effect of GUP on black hole thermodynamics has been studied in other cases as well, (See \cite{Feng1} for Schwarzschild-Tangherlini black hole). The effect of GUP on black hole thermodynamics and phase transition in a cavity has been studied in \cite{Zhou1}. The study of
 generalized Dirac equation and subsequently the thermodynamics due to the tunneling of fermion was studied for  Reissner-Nortdtr\"om black de Sitter quintessence hole in \cite{Feng3}, although the choice of quintessence parameter differs from our work. Similar analysis for Kerr-Newman-AdS black hole was carried out in \cite{Javed}. See also \cite{Tekincay} for the radiation of zitterbewegung particles. All these works use the Hamilton-Jacobi method as opposed to the radial null geodesic approach we adopted. As discussed in Section \ref{sec:label4}, GUP effects do not show up in the tunneling probability when the mass of the black hole is much greater than that of the emitted particle. In this context, we emphasize that only near the Planckian range, when the black hole has evaporated to a remnant limit, does GUP show its noticeable influence. Otherwise, the effect of the GUP parameter is minimal in the plots furnished in Fig. \ref {Fig1}, Fig. \ref{Fig2}, and Fig. \ref{Fig3}. It is worth mentioning that Adler et. al. \cite{Adler} have argued for the existence of a remnant mass black hole as the Planck scale is approached. At
this scale, the black hole stops radiating and its entropy approaches zero, although its effective temperature tries to reach a maximum value. This remnant mass black hole is unable to radiate further, and it
becomes an inert remnant that relies only on gravitational interactions. See also \cite{Carr} in which the existence of a Compton wavelength scale black hole below the Planck mass is predicted.
Fig. \ref{Fig4} shows the energy emission rate from the black hole with respect to the energy of the emitted particle $\Omega$ \cite{Wei} for the massless case. We have plotted the $T_H$ expressions from Table \ref{Table-1} in Fig. \ref{Fig4}. 
\begin{equation}
    \displaystyle\frac{d^2 E(\Omega)}{d\Omega dt}=\displaystyle\frac{2 \pi^2 \sigma_{lim}}{e^{\Omega/T_H}-1}\Omega^3
\end{equation}
Here $ \sigma_{lim}=\pi r_{sh}^2$ is the limiting value of the absorption cross section. It is dependent on the radius of the black hole shadow $r_{sh}$, which we take as $3\sqrt{3}M$ for the Schwarzschild case \cite{Jafarzade}. Compared with \cite{Jamil5}, we see that the figure follows a similar trend. A peak exists for the emission rate in all cases. The peak shifts to a lower frequency range as $w$ from $-2/3$ to $-1/3$, the GUP peak lies somewhere in between.

\section{ Concluding remarks}\label{Conclusion}

\begin{figure}[ht!]
\centering
  \begin{subfigure}[b]{0.475\linewidth}
\centering
\includegraphics[width=\linewidth]{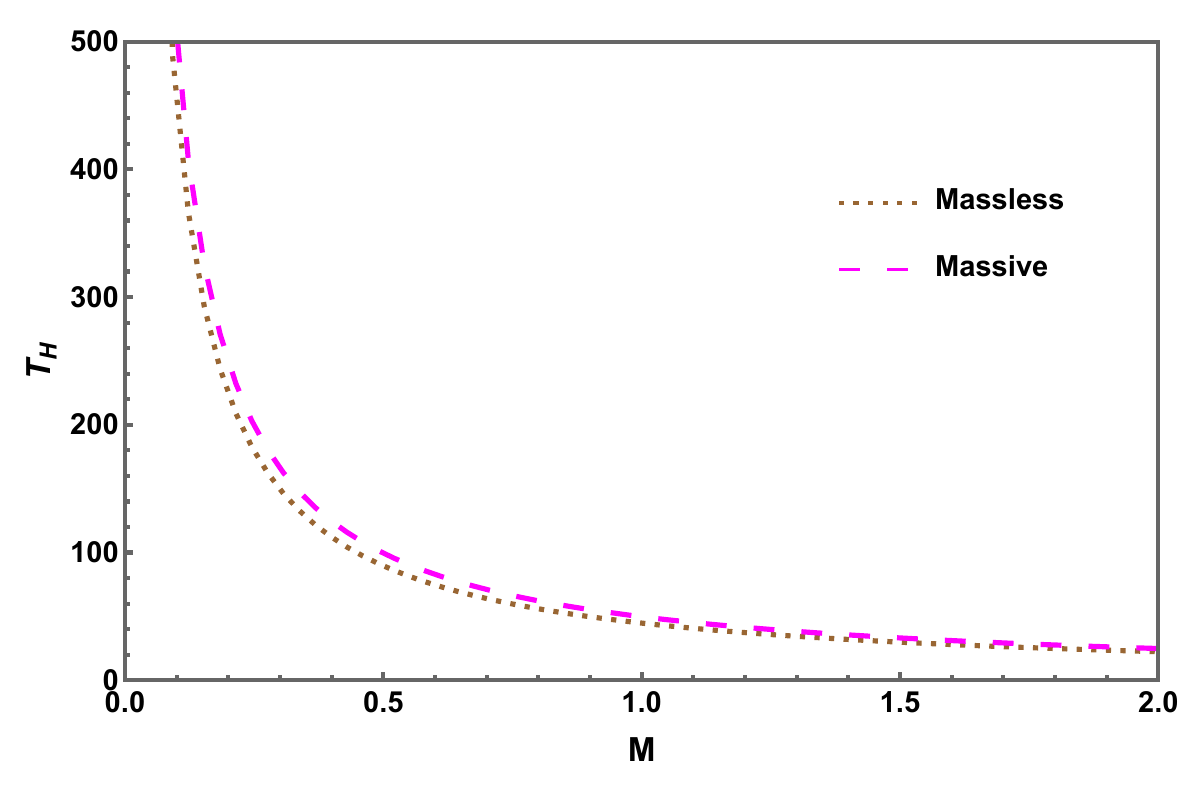}
\caption{}
\label{Fig1a}
\end{subfigure}
\hfill
 \begin{subfigure}[b]{0.475\linewidth}
 \centering
\includegraphics[width=\linewidth]{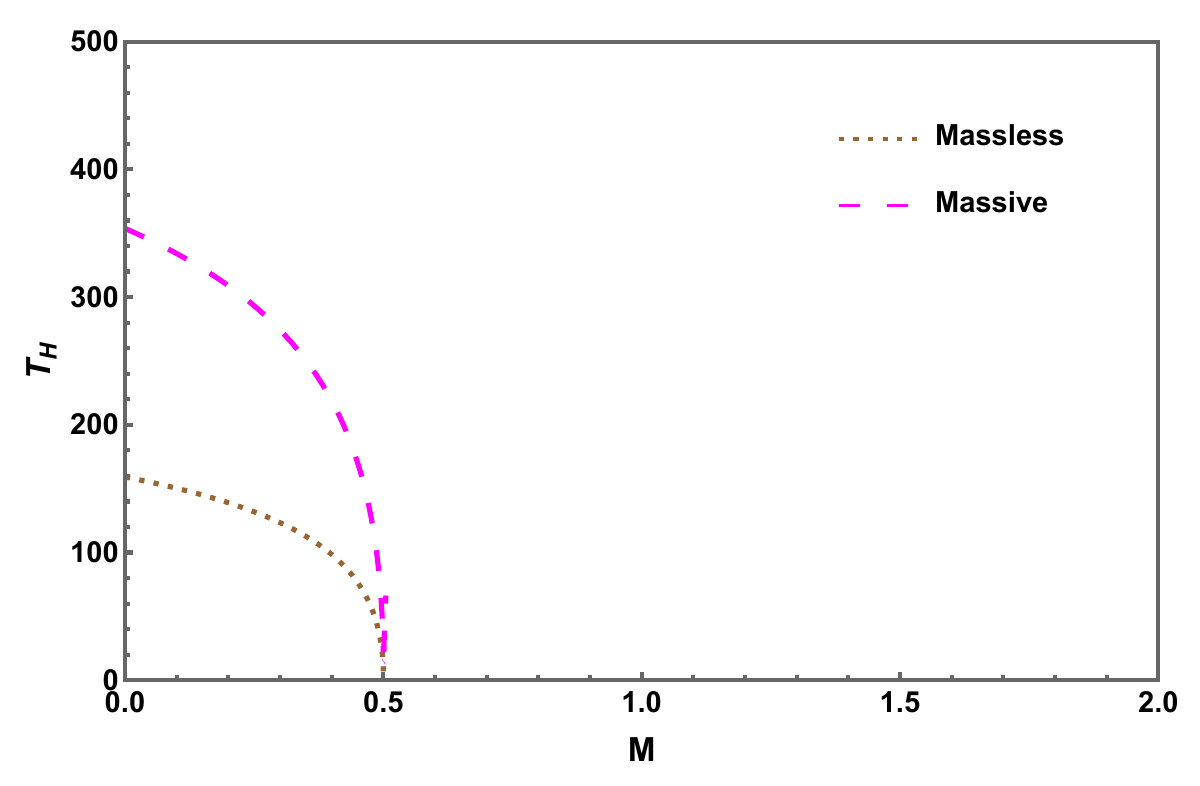} 
\caption{}
\label{Fig1b}
\end{subfigure}

\vspace{0.5em}

\begin{subfigure}[b]{0.475\linewidth}
\centering
\includegraphics[width=\linewidth]{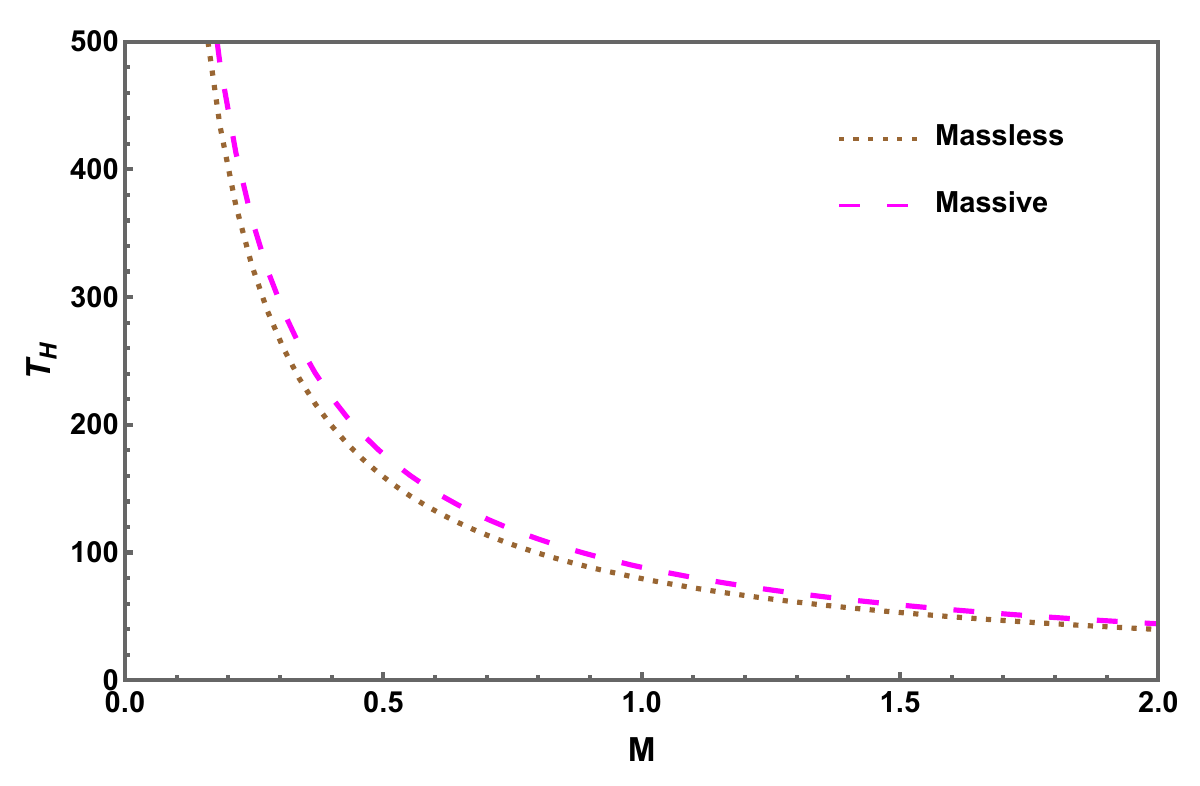}  
\caption{}
\label{Fig1c}
\end{subfigure}

\caption{Hawking temperature $T_H$ vs mass of black hole $M$ for three cases, (\ref{Fig1a}) quintessence case of $w=-1/3$, (\ref{Fig1b}) quintessence case of $w=-2/3$, and (\ref{Fig1c}) GUP case. Comparison of massless vs massive has been done for (\ref{Fig1a}), (\ref{Fig1b}), and (\ref{Fig1c}). For (\ref{Fig1a}) and (\ref{Fig1b}) we have taken $\alpha=0.25$, along with $\Omega=0.001$ and $m=0.0001$. In (\ref{Fig1c}), since $\beta$ is very small, it is neglected. {Temperature for the massless case is always less than the massive case. For $w=-1/3$, the temperature decreases with the increase in mass of the black hole. For $w=-2/3$, because of the condition imposed on $\alpha$, there is an upper bound on the mass of the black hole. $T_H$ and $M$ are dimensionless. See text for more details.}}
\label{Fig1}
\end{figure}

\begin{figure}[ht!]
\centering
\begin{subfigure}[b]{\linewidth}
  \centering
  \includegraphics[width=0.75\linewidth]{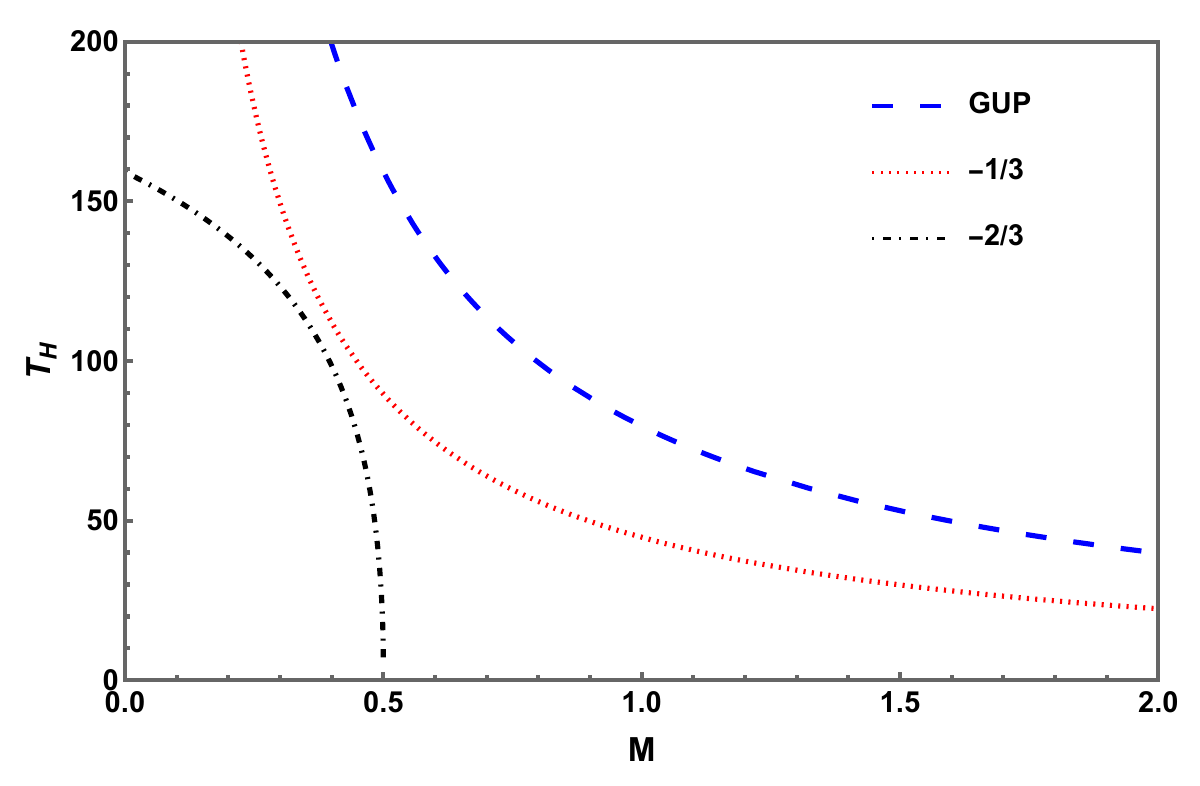}
  \caption{}
  \label{Fig2a}
\end{subfigure}
\begin{subfigure}[b]{\linewidth}
  \centering
  \includegraphics[width=0.75\linewidth]{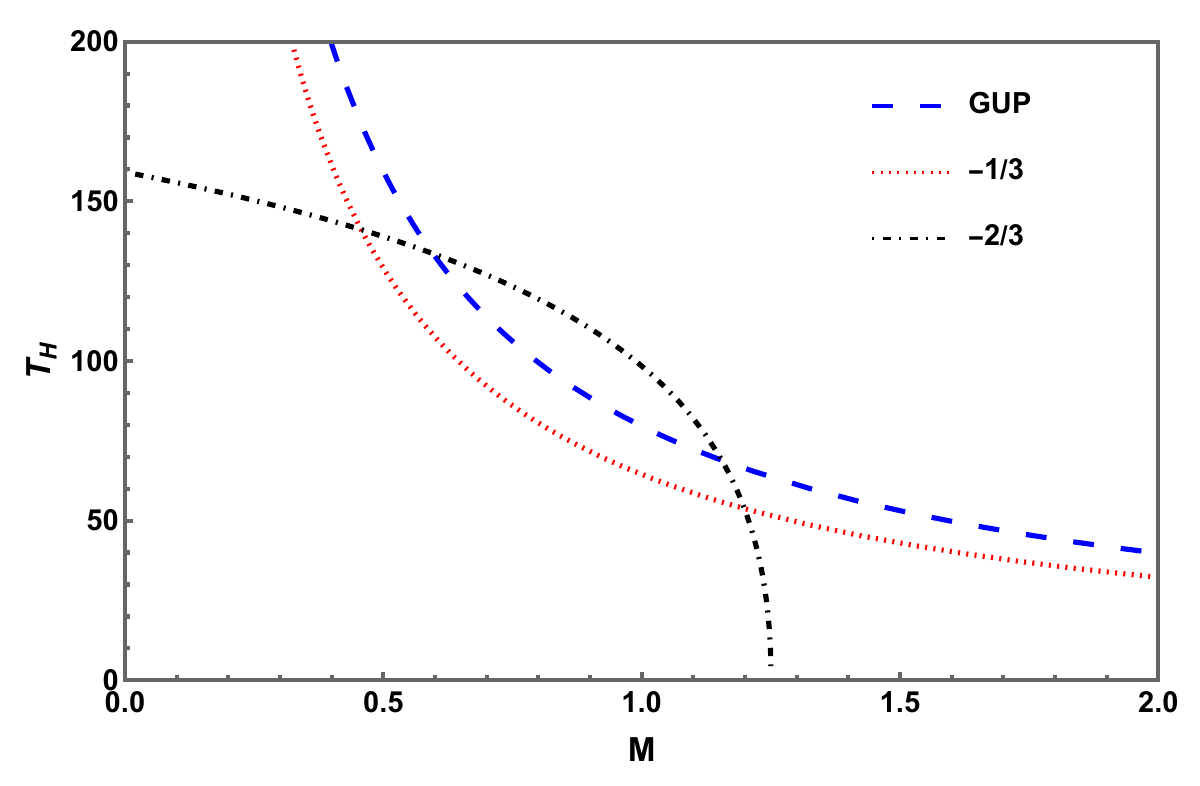} 
\caption{}
 \label{Fig2b}
 \end{subfigure}
 \caption{A comparison of the change in the Hawking temperature $T_H$ as black hole mass $M$ increases for the case of massless particles for GUP, $w=-1/3$, and $w=-2/3$ respectively. The energy $\Omega$ of the emitted massless particle is taken to be $0.001$. The GUP parameter $\beta$ is neglected. (\ref{Fig2a}) $\alpha=0.25$ and (\ref{Fig2b}) $\alpha=0.1$. Non-monotonic trend is observed in both cases. In (\ref{Fig2a}), the temperature is maximum for the GUP case for a given black hole mass, while for $w=-2/3$ it is minimum for the quintessence. This feature is disrupted in (\ref{Fig2b}) in which for $w=-2/3$ the temperature peaks for an intermediate range of $M$. $T_H$ and $M$ are dimensionless. See text for more details.}
\label{Fig2}
\end{figure}

\begin{figure}[ht!]
\centering
\begin{subfigure}[b]{\linewidth}
 \centering
  \includegraphics[width=0.75\linewidth]{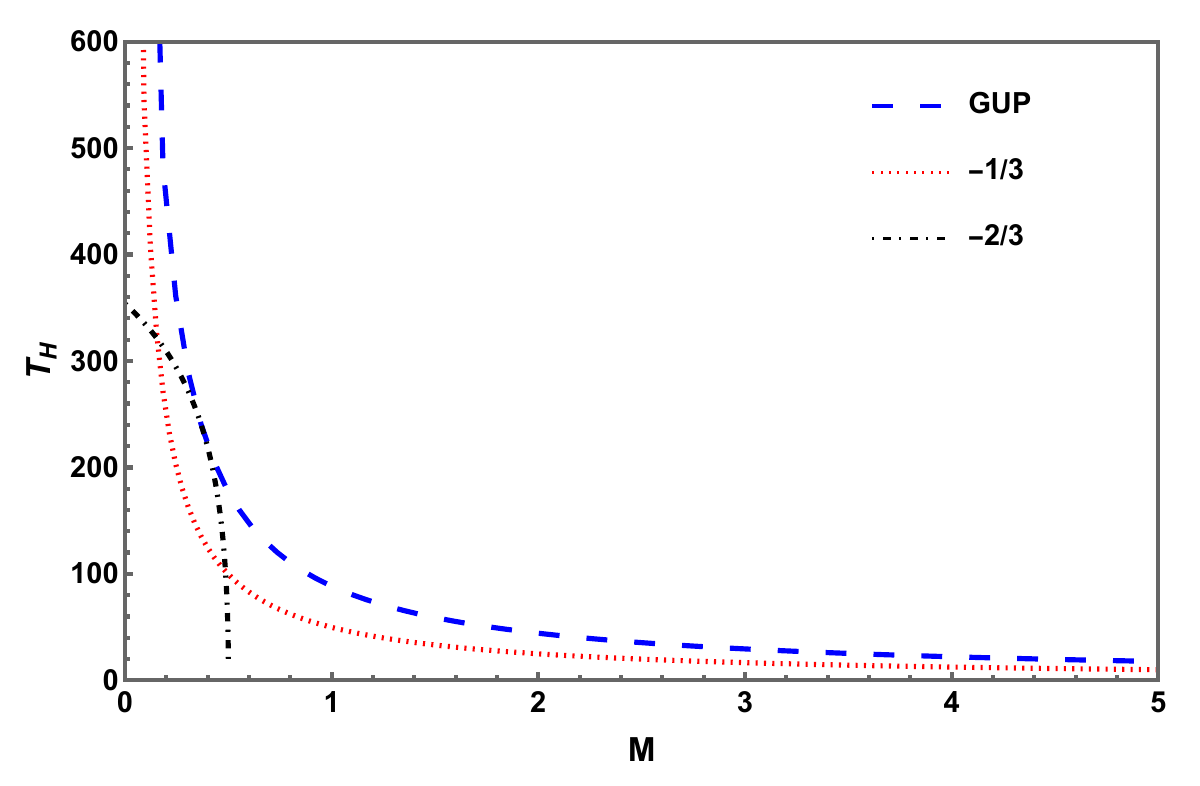}  
  \caption{}
 \label{Fig3a}
\end{subfigure}
 \begin{subfigure}[b]{\linewidth}
\centering
  \includegraphics[width=0.75\linewidth]{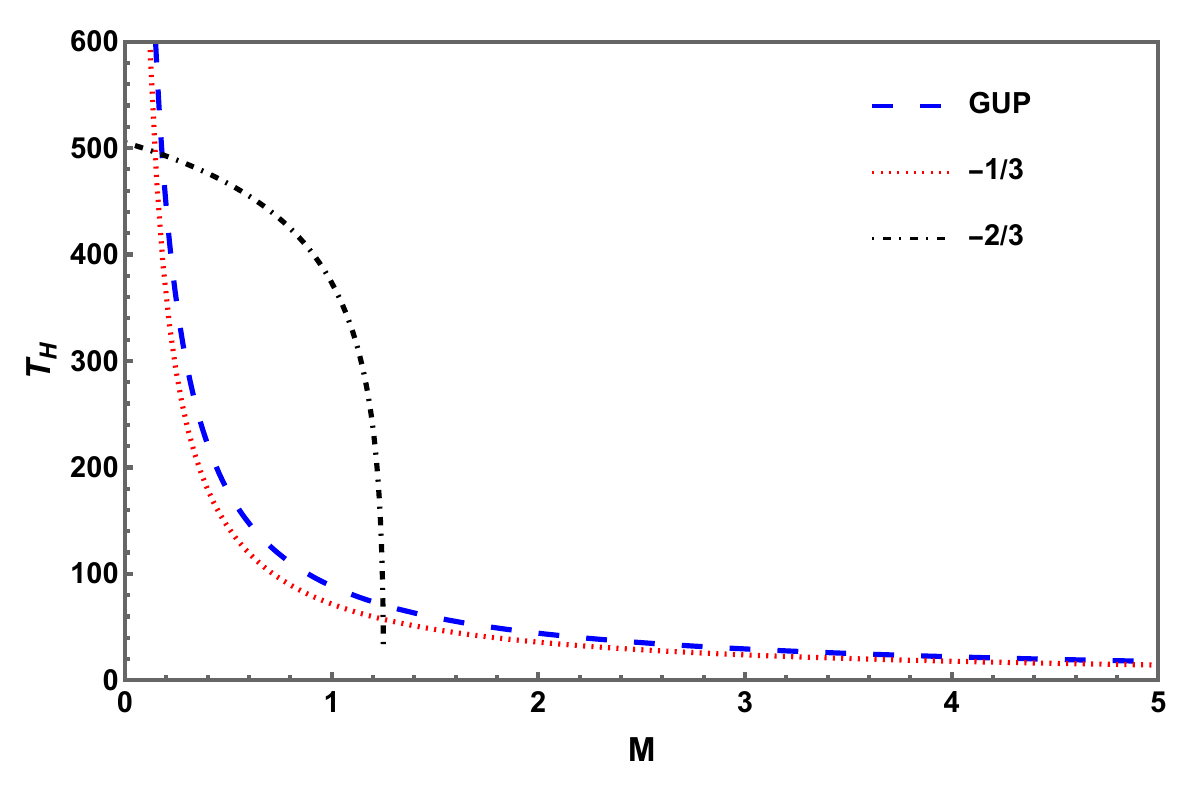} 
  \caption{}
\label{Fig3b}
\end{subfigure}
 \caption{A comparison of the change in the Hawking temperature $T_H$ as black hole mass $M$ increases for the case of massive particles for GUP, $w=-1/3$, and $w=-2/3$ respectively. The energy $\Omega$ of the outgoing particle is taken to be $0.001$, and the lower bound on the energy $m$ is fixed at $0.0001$. The GUP parameter $\beta$ is neglected. (\ref{Fig3a}) $\alpha=0.25$ and (\ref{Fig3b}) $\alpha=0.1$. Similar explanation as Fig. \ref{Fig2}. $T_H$ and $M$ are dimensionless. See text for more details.}
\label{Fig3}
\end{figure}

\begin{figure}[ht!]
\centering
\includegraphics[width=0.7\linewidth]{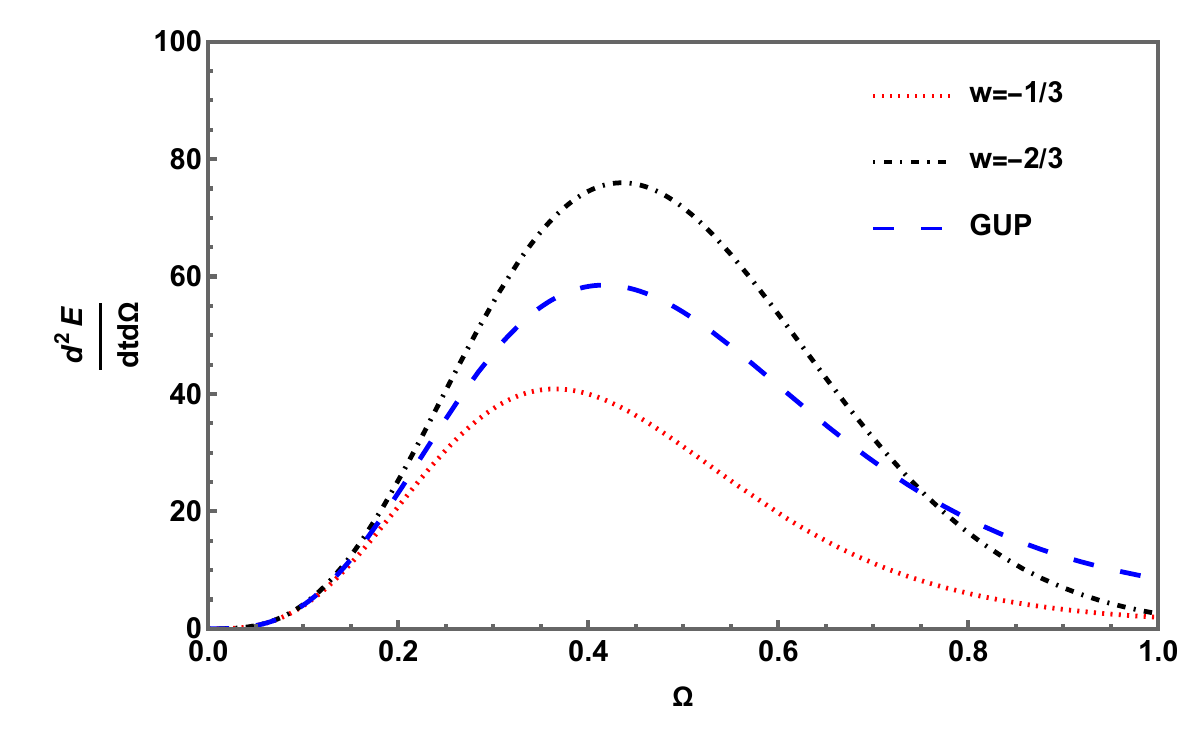}  
\caption{Energy emission rate $\frac{d^2E}{d\Omega dt}$ vs energy of emitted particle $\Omega$ for massless particle. Both the energy emission rate and the energy of the emitted particle are dimensionless. Here $\alpha=0.1$, the mass of the black hole $M$ is taken as unity. Shadow radius is $r_{sh}=3\sqrt{3}M$. The GUP parameter $\beta$ is neglected. $T_H$ has been taken from Table \ref{Table-1}. See text for more details.}
\label{Fig4}
\end{figure}

In this work, we have reported the tunneling mechanism and some thermodynamic aspects of a modified Schwarzschild black hole in the context of quintessence and the GUP principle. Our aim has been to revisit the problem of tunneling again for completeness with the intention of tackling the contour integration with some rigor. The novelty of our work lies in the quarter-circle contour of integration, which we introduced for calculating the tunneling probability. The goal was to compare the effects of a classical and quantum phenomenon on the tunneling probability in an attempt to better understand the semiclassical approach of tunneling. As already mentioned, an important aspect of our work is the employment of a unique contour of integration to calculate the tunneling probability. We demonstrated that the use of a quarter-circle contour around the upper limit of the concerned integration facilitated a straightforward determination of the latter. For such a choice of the contour, we were motivated in spirit from the works of Stotyn \cite{Stotyn} and also Kerner and Mann \cite{Mann3}. We calculated the Hawking temperature for both the cases of massless and massive particles and observed from Fig. \ref{Fig1} that the temperature values in the massless case are always less than the massive case. In particular, for the $w=-1/3$ case of the quintessence, the temperature curve rapidly decreases with the increase in mass of the black hole. On the other hand, the $w=-2/3$ case of the quintessence, because of the condition imposed on $\alpha$, reveals an upper bound on the mass of the black hole. We also found from our analysis that the tunneling probability of the emitted particle is always higher for the massive case than its massless counterpart for both quintessence and GUP. It needs to be pointed out that the analysis done in \cite{Lutfuouglu} differs from our work in that while it studies the GUP effects on the thermodynamics of the Schwarzschild black hole, which is already surrounded by quintessence, we have highlighted the contrasting picture of both corrections (quintessence and GUP). We emphasize that our works support the presence of a remnant mass and remnant temperature, thereby setting a lower bound on $M$ while the black hole undergoes an evaporation (in this regard, see also \cite{Sunandan1} which focuses on temperature and entropy corrections). 
\footnote{Note that Table-\ref{Table-1} suggests that the temperature is unphysical in the limit $M \rightarrow 0$, even if it provides a non-zero number for the quintessence scenario. It would imply that the energy of the released particle $\Omega$ determines the temperature of the fully evaporated black hole. This leads to an inconsistency. Thus, it is clear that leftover mass exists. Therefore, it is possible to state a lower constraint on the black hole mass $M_{min}$, where $M_{min}>\Omega$ is always true. The GUP scenario, when the logarithmic function blows out due to the $M\rightarrow0$ limit, may be analyzed similarly. In accordance with the explanation provided prior to equation (\ref{eq47}), the presence of $M_{min}$ eventually places a strict upper restriction on the value of $\alpha$.}
\footnote{A comparative investigation of the identical tunneling process caused by two radically different regimes, classical and quantum, is provided by the parameters $\alpha$ and $\beta$ employed in our study. In \Cref{Fig1,Fig2,Fig3}, we ultimately disregard the $\beta$ parameter because it has a negligible influence and doesn't alter the traditional result in any way. Therefore, we find that, for $M>>\Omega$, the GUP adjustment to the Schwarzschild metric may be ignored.} We have also analyzed the energy emission rate of the black hole with respect to the energy of the emitted particle and found that the trend is similar to what is found in the literature. A peak exists for the emission rate in all cases.

 Finally, an interesting comparison can be drawn between \cite{Xu} and our work, where the existence of the state parameter in the case of a rotating Kiselev black hole is observed. In the non-rotating limit, in their work, for the two cases of $w=-1/3$ and $w=-2/3$, the result simplifies down to functions of $M$ and $\alpha$. The variation in the exact form arises from the definition of the metric element for the rotating case. It is worth noting that some recent works have tried to put constraints on the GUP parameter $\beta$ using astrophysical observations of black hole shadows, lensing and quasiperiodic oscillations \cite{Jamil1,Jamil2,Jamil3,Jamil4}.
 It is well known that the spin of the outgoing particle also affects the tunneling probability. Spin$-0$ particles evolve using the Klein-Gordon equation, whereas the spin$-1/2$ particles follow the Dirac equation. Tunneling for both such particles has been extensively studied in the literature \cite{Mann1,Mann2,Mann3,Majhi1}. \footnote{ Also see \cite{Li, Yang, Yale}, which explores the broader applicability of the tunneling mechanism in the case of fermions and bosons. We remark that these works focus on the Hamilton-Jacobi formalism of tunneling, whereas our current work took the null geodesic approach. Tunneling of such spin$-0$ and spin$-1/2$ particles and the effect of quintessence and GUP on their tunneling probability would be explored in future projects.}\footnote{It would be interesting to study the tunneling of neutrinos and gravitons in the case of quintessence background and GUP. See \cite{Yang} for neutrino tunneling from NUT Kerr Newman de Sitter black hole. \cite{deVries} presents that tunneling of neutrinos, electrons, and gravitons is only possible in an extremally rotating Kerr black hole. }

\section{Acknowledgements}
The author thanks Bijan Bagchi for guidance. He also thanks Rahul Ghosh for insightful discussions. Encouraging communications from Sara Saghafi and E. C. Vagenas are gratefully acknowledged. The author acknowledges financial support from the Shiv Nadar Institution of Eminence, and the Council of Scientific and Industrial Research (CSIR), Government of India for a direct SRF fellowship under grant 09/1128(18274)/2024-EMR-I. The author is also grateful to the anonymous reviewers for constructive criticism and for providing useful suggestions.
\begingroup

\setlength{\tabcolsep}{10pt} 
\renewcommand{\arraystretch}{2} 

\begin{table*}
\caption{We summarize Hawking temperature for the various cases of tunneling, for both massless and massive particles. The first row is subdivided further into two cases of quintessence corresponding to the values of $w=-1/3$ and $w=-2/3$. The second row stands for the GUP case. }
\renewcommand{\arraystretch}{2}
\vspace{2mm}
\centering
  \resizebox{\textwidth}{!}{%
\begin{tabular}{|cl|l|l|}
\hline
\hline
\multicolumn{2}{|c|}{\textbf{Quintessence/GUP}}                         & \multicolumn{1}{c|}{\textbf{$T_H$ for massless particles}} & \multicolumn{1}{c|}{\textbf{$T_H$ for massive particles}} \\ 
\hline
\hline
\multicolumn{1}{|c|}{\multirow{2}{*}{Quintessence}} & \multicolumn{1}{c|}{$w=-1/3$} &   \multicolumn{1}{c|}{$\displaystyle\frac{(1-\alpha)^2}{4\pi\left( M\Omega - \frac{\Omega^2}{2} \right)}$  }          &     \multicolumn{1}{c|}{$\displaystyle\frac{(1-\alpha)^2}{4\pi\left( M\Omega - \frac{\Omega^2}{2} -Mm +\frac{m^2}{2}\right)}$  }                       \\  
\multicolumn{1}{|c|}{}                              & \multicolumn{1}{c|}{$w=-2/3$} & \multicolumn{1}{c|}{$\displaystyle\frac{1}{\pi\left( \Omega + \frac{1}{4\alpha}\left(\sqrt{1-8\alpha(M-\Omega)}-\sqrt{1-8\alpha M}\right)    \right)}$}     &    \multicolumn{1}{c|}{$\displaystyle\frac{1}{\pi\left( \frac{(1+\alpha m^2)}{2} \left\{ (\Omega-m) + \frac{1}{4\alpha}\left( \sqrt{1-8\alpha(M-\Omega)} - \sqrt{1-8\alpha(M-m)} \right) \right\} \right)}$}           \\ \hline
\multicolumn{2}{|c|}{GUP}          &    \multicolumn{1}{c|}{$\displaystyle\frac{1}{2\pi \left( (2M-4\beta)\Omega - \Omega^2 -8\beta \ln \left( \frac{|\Omega-M|}{M} \right)  \right)}$}                     &         \multicolumn{1}{c|}{$\displaystyle\frac{1}{2\pi \left(  (2M-4\beta) (\Omega-m) - (\Omega^2-m^2)-8\beta \ln \left( \frac{|\Omega-M|}{|m-M|}  \right) \right)}$}     \\ \hline
\end{tabular}%
}
\label{Table-1}
\end{table*}
\endgroup

\appendix

\section{Solution of the radial part of the integral (\ref{Omega}) for the massless case}\label{Appendix-A}

Let us take the integral (\ref{Omega}) for the $w=-1/3$ massless case where $f^{(q)}(r)$ can be deduced from (\ref{eq5.a.1.1}). The radii $r_{in}$ and $r_{out}$ are respectively, $\frac{2M}{1-\alpha}$ and $\frac{2(M-\Omega)}{1-\alpha}$. Since $f^{(q)}(r)$ is small near the horizon, we expand it binomially and proceed according to \cite{LandauQM,Srini_Paddy} to obtain
\begin{eqnarray}
     \mbox{Im}\zeta &=& 2\;\mbox{Im} \int_{r_{in}}^{r_{out}} \int_0^\Omega \frac{d\Omega}{f^{(q)}(r)} dr \nonumber\\
&=& \frac{2}{1-\alpha} \mbox{Im} \int_{r_{in}}^{r_{out}} \int_0^\Omega \frac{r}{r-\frac{2(M-\Omega)}{1-\alpha}}d\Omega\, dr,  \end{eqnarray}
where observe that there is a pole in the upper limit at $r_{out}$. Changing the order of integration we rewrite the integral for $\mbox{Im}\, \zeta$ in the manner
\begin{eqnarray}
    \mbox{Im}\, \zeta &=& \frac{2}{1-\alpha} \mbox{Im} \int_{r_{in}}^{r_{out}} \int_0^\Omega \frac{r}{r-\frac{2(M-\Omega)}{1-\alpha}} \, d\Omega \, dr \nonumber \\
    &=& \frac{2}{1-\alpha} \mbox{Im} \int_{r_{in}}^{r_{out}} \int_0^\Omega \frac{r}{r-r_{out}} \, d\Omega \, dr. 
\end{eqnarray}

\begin{figure}[ht!]
    \centering    \includegraphics[scale=0.15]{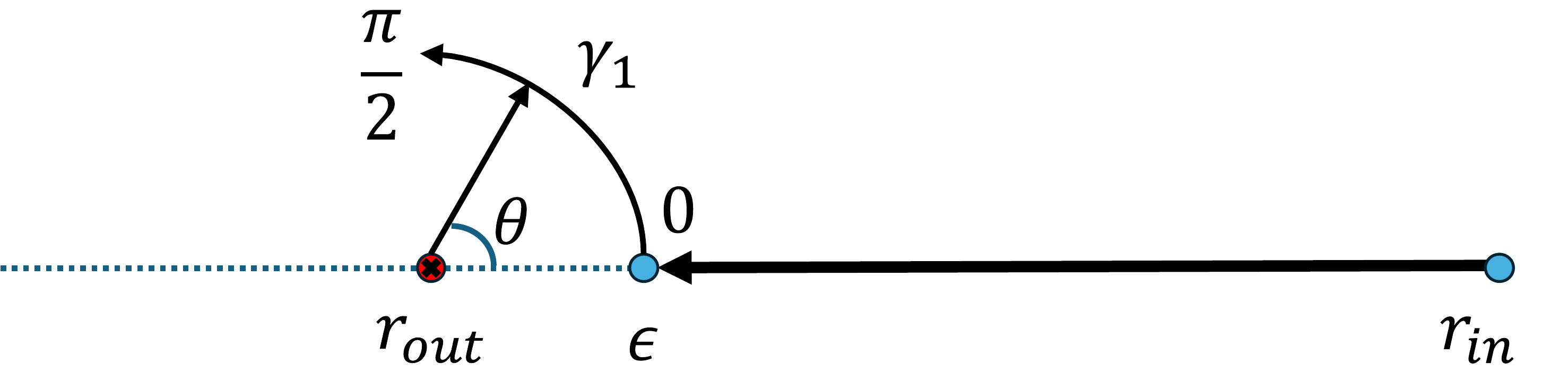}
     \caption{(Color online) Contour of integration for Appendix \ref{Appendix-A}.}
     \label{contour1}
\end{figure}

Breaking the integration into two parts and adopting a contour described in Fig. \ref{contour1}, wherein the quarter circle contour $\gamma_1$ runs counterclockwise from $0$ to $\pi/2$, we transform  $r \rightarrow r_{out}+\epsilon e^{i \theta}$ to obtain
  
\begin{equation}
    \mbox{Im}\, \zeta = \frac{2}{1-\alpha} \mbox{Im} \int_0^\Omega \left[\lim_{\epsilon\to 0} \left( \cancelto{Real}{\int_{r_{in}}^{\epsilon}} + \int_0^{\pi/2}  \right) \frac{r_{out}+\epsilon e^{i\theta}}{\cancel{r_{out}}+\epsilon e^{i\theta}-\cancel{r_{out}}}i\epsilon e^{i\theta} d\theta \right] d\Omega.
\end{equation}

The first integral doesn't contribute any imaginary term, which is needed for the tunneling probability. Switching the limits for the second integral, the imaginary action turns out to be 
\begin{eqnarray}
    \mbox{Im} \zeta &=& \frac{2}{1-\alpha} \frac{\pi}{2}\int_0^\Omega r_{out} \;d\Omega \nonumber \\
    &=& \frac{2\pi}{(1-\alpha)^2} \left( M\Omega - \frac{\Omega^2}{2} \right).
\end{eqnarray}

\section{Solution of the radial part of the integral (\ref{eq5.13}) for massive case}\label{Appendix-B}

The radial integral (\ref{eq5.13}) in the massive case consists of two integrals $I_1$ and $I_2$.
As an illustration, we solve them for the quintessence case corresponding to the ratio $w=-1/3$, noting that an analogous treatment holds for $w=-2/3$. One can also handle the GUP case in a similar way.

\subsection{Solution of integral (\ref{eq5.15a})}

\begin{figure}[ht!]
    \centering    \includegraphics[scale=0.15]{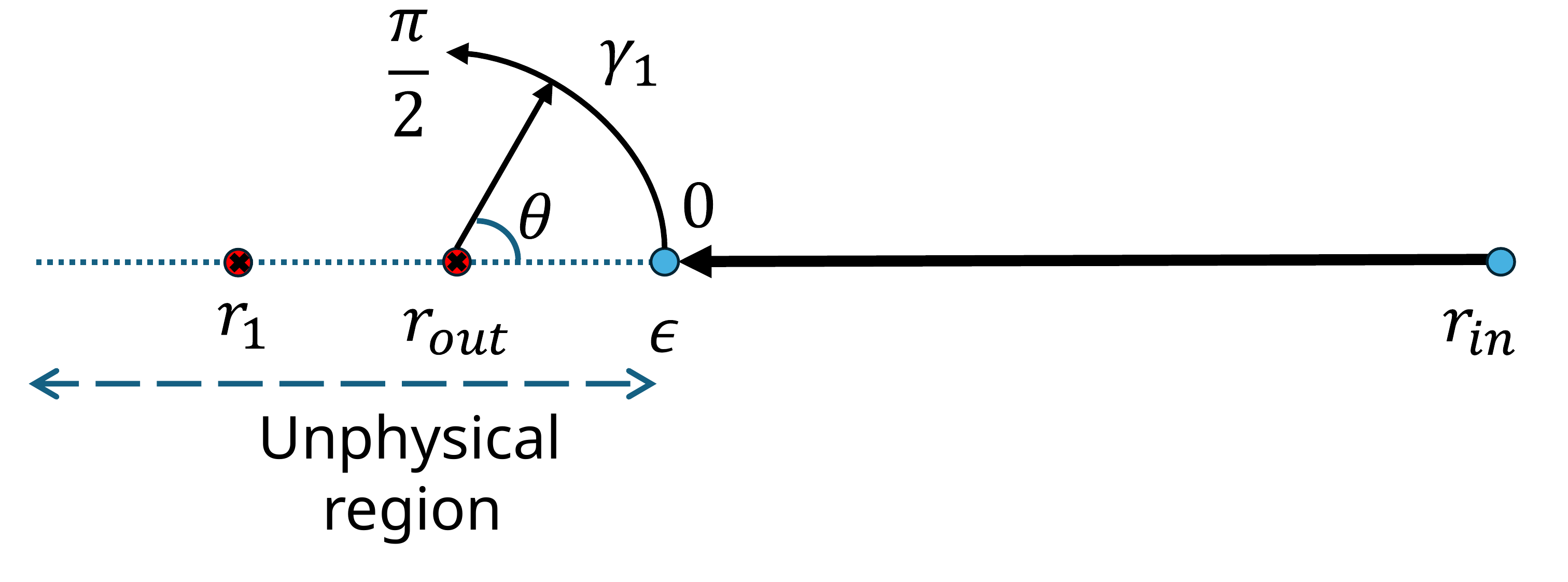}
     \caption{(Color online) Contour of integration for Appendix-\ref{Appendix-B}.} 
         \label{contour2}
\end{figure}

The integral $I_1$ can be reduced to the form 
\begin{equation}\label{eq5.a.1.4}
    I_1 = \varrho\int_{r_{in}}^{r_{out}} \frac{r^2}{(r-r_{out})(r-r_1)}dr,
\end{equation}
where the upper limit signifies the location of one of the poles at 
$r_{out}$. Here $r_1 = -\frac{2m^2(M-\Omega)}{2\Omega^2-m^2(1-\alpha)}$ and $\varrho=\frac{2\Omega^2}{(1-\alpha)(2\Omega^2-m^2(1-\alpha))}$. Of course, the other pole at $r_1$ lies in the unphysical region.
The contour of integration is described in Fig. \ref{contour2} where the integration runs over the real line from $r_{in}$ to $\epsilon$, along with a quarter circle $\gamma_1$, in which the angle $\theta$ varies from $0$ to $\pi/2$, to avoid the pole at $r_{out}$. The integral $I_1$ can then be represented as
\begin{equation}\label{eq5.a.1.5}
I_1 =  \varrho\left[\lim_{\epsilon\to0} \left( \cancelto{Real}{\int_{r_{in}}^{\epsilon}} + \int_{\gamma_1}  \right) \frac{(r_{out}+\epsilon e^{i\theta})^2}{(\cancel{r_{out}}+\epsilon e^{i\theta}-\cancel{r_{out}})(r_{out}+\epsilon e^{i\theta}-r_1)}i\epsilon e^{i\theta} d\theta \right],
\end{equation}
where the domain of the first integral is over the real region and hence does not contribute to the tunneling probability. For the second integral, we notice that it is bounded from above, and hence we can switch the order of the integral to express
\begin{eqnarray}\label{eq5.a.1.6}
I_1 &=& \varrho \int_o^{\pi/2} \lim_{\epsilon\to0} \frac{r_{out}^2+\epsilon^2 e^{2 i \theta}+2 \epsilon r_{out} e^{i \theta}}{r_{out}-r_1+\epsilon e^{i\theta}} i d\theta \nonumber\\
&=& i\frac{\pi}{2} \varrho \frac{r_{out}^2}{(r_{out}-r_1)} \nonumber \\
&=& \frac{i\pi}{(1-\alpha)^2}(M-\Omega).
\end{eqnarray}

\subsection{Solution of integral (\ref{eq5.15b})}

The integral $I_2$ can be cast in the form
\begin{equation}\label{eq5.a.1.7} 
I_2\approx  \int_{r_{in}}^{r_{out}} \frac{1}{f^{(q)}(r)}dr - \cancelto{Real}{\int_{r_{in}}^{r_{out}}\frac{1}{2}dr} ,   
\end{equation}
where the second integral is discarded because it only gives a real value. For the first integral, we find that there is only one pole at $r_{out}$ for which we get by proceeding similarly as done for the integral $I_1$ the estimate
\begin{equation}
    I_2 = \frac{i\pi}{(1-\alpha)^2}(M-\Omega).
\end{equation}
With the values of $I_1$ and $I_2$ at hand we determine from (\ref{eq5.14}) the following evaluation of $\mbox{Im}\,\zeta$
\begin{eqnarray}
\mbox{Im}\,\zeta &=& \frac{2\pi}{(1-\alpha)^2}\int_m^\Omega (M-\Omega)\, d\Omega  \nonumber\\
&=& \frac{2\pi}{(1-\alpha)^2}\left(M\Omega-\frac{\Omega^2}{2}-mM+\frac{m^2}{2}\right).
\end{eqnarray}

\bibliographystyle{ieeetr}
\bibliography{biblio.bib}


\end{document}